\documentclass[final]{ias2}

\usepackage{graphicx}
\usepackage{multirow}
\usepackage{array}

\usepackage{hyperref}
\def\be{\begin{equation}}
\def\ee{\end{equation}}
\def\ba{\begin{array}}
\def\ea{\end{array}}
\def\bea{\begin{eqnarray}}
\def\eea{\end{eqnarray}}
\begin{document}


\title{Properties of N, $\Delta$ Baryons with Screened Potential}

\author[sin]{C\lowercase{handni} M\lowercase{enapara}}
\email{chandni.menapara@gmail.com}
\author[sin]{A\lowercase{jay} K\lowercase{umar} R\lowercase{ai}}
\address[sin]{Department of Physics, Sardar Vallabhbhai National Institute of Technology, Surat-395007, Gujarat, India}

\begin{abstract}
N and $\Delta$ baryons hold an important place towards understanding the quark dynamics inside hadrons. The hypercentral Constituent Quark Model (hCQM) has been employed in various studies ranging from light to heavy hadrons. In the present article, screened potential has been used to study light baryon resonances. The Regge trajectories have been plotted alongwith the details of slopes and intercepts. The strong decay widths to pion have been calculated for some channels using the present masses.

\end{abstract}



\maketitle

\section{Introduction}
Hadron spectroscopy is an important tool to understand the quark dynamics inside hadrons \cite{crede, thiel, eichmann}. Most nuclear phenomena can be explained in terms of the non-relativistic interactions between protons and neutrons, which make up the nucleus's elementary constituents. Quantum Chromodynamics (QCD), on the other hand, is the theory behind nuclear forces and describes relativistic quarks and gluons as the fundamental degrees of freedom \cite{qcd}. Confinement serves as QCD's defining characteristic. It appears to prohibit the free existence of isolated elementary quarks and gluons in nature. One of the central issues in physics is how to understand how confinement arises. Confinement results from the self-interactions of gluons, which act as the strong force's intermediaries between colored quarks and other gluons. Since the light, u, and d quarks that make up the nuclei are many times lighter than the proton, the majority of the visible mass in the universe is actually produced by relativistic interactions between gluons. \\

$N^{*}$ corresponding to P(uud) and N(udd) with $J=\frac{1}{2}$ and isospin $I=\frac{1}{2}$ has always been in discussion for decades \cite{pdg}. N. Isgur has nicely highlighted few facts as to why to study of $N^{*}$ have always been in the priority as: The whole stable matter around us is made up of nucleons; it being the simplest system to manifest the non-abelian nature of QCD. However, inspite of being the lowest state, it has been known through understanding of years, that hadrons are a really complex systems whose all the properties are not known to us till date.  \\

Over the years, many approaches have been implemented with the intention to understand thoroughly the light baryon sector. Recently all light and strange baryons have been studied through Bethe Ansatz method with U(7) by an algebraic method \cite{amiri}. The earlier algebraic method has been discussed by R. Bijker et.al. to study baryon resonances in terms of string-like model \cite{bijker}. A. V. Anisovich et al. has reproduced the N and $\Delta$ spectrum using the multichannel partial wave analysis of pion and photo-induced reactions \cite{thoma}. The quark-diquark model using Gursey Radicati-inspired exchange interaction has been studied \cite{s5,s15}. The semi-relativistic constituent quark model, classification number describing baryon mass range \cite{chen}, mass formula obtained by Klempt \cite{klempt}, dynamical chirally improved quarks by BGR \cite{bgr} are among various models. Also, the relativistic study with quark-diquark model have been studied for all sectors by Faustov et al. \cite{galkin}. A recent study based on HAL-QCD, decuplet baryons have been focused wherein interaction potentials extracted from lattice QCD \cite{hal-qcd}. Regge phenomenology has also been employed in the study of light, strange baryons using n and J plane linear curves \cite{juhi}.\\

The present study is based on a non-relativistic approach namely hypercentral Constituent Quark Model (hCQM). Our earlier works deal with hCQM applied for hadrons from light to heavy sector. The screened potential term is accompanied with spin-dependent term in the present approach. The section 3 discusses mass spectra of N and $\Delta$ baryons alongwith the experimental and theoretical background known so far. Section 4 is dedicated to Regge trajectory for (J,$M^2$) and (n,$M^2$). The strong decay channels with pion have been studied in section 5.

\section{Theoretical Framework}

The choice of a hypercentral SU(6) invariant potential, sometimes referred to as a potential whose value is exclusively governed by the hyperradius x, is the foundation of the hypercentral Constituent Quark Model, abbreviated as hCQM. In addition to making the process of finding a solution for the Schrodinger equation more straightforward, the selection of a hypercentral potential also has some intriguing ramifications in terms of the physical world.
As a result of the fact that the hyperradius x is dependent on each of the three constituent coordinates at the same time, a hypercentral potential is not just an interaction between two bodies but can also involve terms involving three bodies \cite{gianinni1,gianinni}. hCQM has been applied in various systems and with a variety of potentials by our team for heavy hadrons \cite{zalak19,zalak18,zalak16,zalak17,keval18,keval20,amee1,amee2,amee3} and exotics. The linear potential has been employed for all octet, decuplet baryons in our earlier works \cite{cpc1,cpc2,cpc3,ijmpa1,ijmpa2}.\\

Because of the non-abelian nature of QCD, which results in gluon-gluon interactions, which, in turn, can produce three body forces, these terms have the potential to play an important part in the description of hadrons.
The space component of the 3q wave function, on the other hand, can be expanded in the hyperspherical harmonics basis, at which point the Schrodinger equation transforms into a set of coupled differential equations.
Additionally, it has been shown that the low lying resonance states can be adequately described by the hyperspherical approximation, which is the assumption that only the first term in the expansion of the potential in terms of hyperspherical harmonics is maintained.
Keeping these considerations in mind, a hypercentral potential can be interpreted in one of two ways: either as a standard two body potential or as a three body potential, both of which are treated in the hypercentral approximation.
First of all, we introduce hyperspherical coordinates as hyperradius x and hyperangle $\xi$ from Jacobi coordinates as \cite{bijker},
\begin{equation}
	x = \sqrt{{\bf \rho^{2}} + {\bf \lambda^{2}}} ; \; \; \xi = arctan(\frac{\rho}{\lambda})
\end{equation}
The model itself suggets that the potential to be chosen should be hypercentral i.e. depending only on hyperradius x. The hyperradius x depends at the same time on all the three constituent coordinates, therefore an hypercentral potential is not a pure two body interaction but can also contain three body terms.
These terms can play an important role in the description of hadrons, since the non-abelian nature of QCD leads to gluon-gluon interaction which can, in turn, produce three body forces. On the other hand the space part of the 3q wave function can be expanded in the hyperspherical harmonics basis and the Schrodinger equation becomes a set of coupled differential equations.
Thus, the spatial wave-function can be expressed as hyper-radial part and hyperspherical harmonics as
\begin{equation}
	\psi_{space}= \psi(x) Y(\Omega_{\rho},\Omega_{\lambda},\xi)
\end{equation}  
\begin{equation}
	L^2Y_{[\gamma]}l_{\rho}l_{\lambda}(\Omega_{\rho}, \Omega_{\lambda}, \xi)=-\gamma(\gamma+4)Y_{[\gamma]}l_{\rho}l_{\lambda}(\Omega_{\rho}, \Omega_{\lambda}, \xi)
\end{equation}
where, $\Omega_{\rho}$ and $\Omega_{\lambda}$ are the angle of hyperspherical coordinates. $\vec{L}=\vec{L_{\rho}}+\vec{L_{\lambda}}$ is the total angular momentum and $l_{\rho}$ and $l_{\lambda}$ are the angular momenta associated with the Jacobi coordinates $\rho$ and $\lambda$ respectively. $-\gamma(\gamma+4)$ gives the eigenvalues of $L^2$, where, $\gamma=2n+\rho+\lambda$ is the grand angular momentum quantum number which acquires positive integer value.

The hyper-radial equation whose solution is $\psi(x)$ is as follows,
\begin{equation}
	\left[\frac{d^{2}}{dx^{2}} + \frac{5}{x}\frac{d}{dx} - \frac{\gamma(\gamma +4)}{x^{2}}\right]\psi(x) = -2m[E-V_{3q}(x)]\psi(x) 
\end{equation}
Another choice of hypercentral potential has brought us to screened potential of the form as described below \cite{vijande}. 
\begin{equation}
	V^{0}(x)=a\left(\frac{1-e^{-{\mu} x}}{\mu}\right)
\end{equation}
Such potential has been known to show good results using hCQM for heavy quark system including mesons and baryons \cite{wang}. However, here we have attempted to see such effect in light, strange systems as well. The screening parameter is different in case of heavy and light systems. Also, the results have been discussed for this study here. Based on a paper by R. Chaturvedi, the screening parameter $\mu$ has been varied over a range and 0.3 has been considered as the value obtain the spectra for all the systems considered here.\cite{raghav} \\
Also, this potential form cannot account for splittings of multiplet levels. So, an additional hyperfine splitting term is to be incorporated. However, the spin-dependent interaction $V_{SD}(x)$ consists of three types of interactions, i.e. spin-orbit, spin-spin and tensor terms \cite{voloshin}.
\begin{eqnarray}
	V_{SD}(x) = V_{LS}(x)({\bf L \cdot S}) + V_{SS}(x) \left[S(S+1)-\frac{3}{2}\right] \\ + V_{T}(x)\left[S(S+1) -\frac{3({\bf S \cdot x})({\bf S \cdot x})}{x^{2}} \right] 
\end{eqnarray}
Let, $V_{V}= \frac{\tau}{x} $ and $V_{S}= \alpha x $ \\
The spin-orbit term, 
\begin{equation}
 V_{LS}(x)= \frac{1}{2m_{\rho}m_{\lambda}x}\left( 3\frac{dV_{V}}{dx} - \frac{dV_{S}}{dx} \right) 
\end{equation}
The spin-spin term,
\begin{equation}
	V_{SS}(x) = \frac{1}{3m_{\rho}m_{\lambda}} \nabla^{2}V_{V}
\end{equation} 
The tensor term,
\begin{equation}
V_{T}(x) = \frac{1}{6m_{\rho}m_{\lambda}} \left( \frac{d^{2}V_{V}}{dx^{2}}- \frac{1}{x} \frac{dV_{V}}{dx} \right) 
\end{equation}
{\bf L} and {\bf S} represent the angular momentum and total spin. 

\begin{equation}
	\overrightarrow{s_1} \cdot \overrightarrow{s_2}=\frac{1}{2}\left[\vec{S}^2-s_1\left(s_1+1\right)-s_2\left(s_2+1\right)\right] 
\end{equation}
\begin{equation}
	\vec{L} \cdot \vec{S}=\frac{1}{2}[j(j+1)-l(l+1)-S(S+1)]
\end{equation}
\begin{equation}
	S_{12}=2\left[3 \frac{(\vec{S} \cdot \vec{r})^2}{r^2}-\vec{S}^2\right]
\end{equation}
Here, $s_1$ and $s_2$ denote the spins of individual quarks inside the baryon. L and S denote the total orbital angular momentum and spin quantum numbers for a given state.
All these have been carried out with Mathematica notebook \cite{lucha}.
\section{Results and discussions}

The latest update of Particle Data Group (PDG) lists the most recent and precise ground state mass to be $m_{p}= 938.272 MeV $ and $m_{N}= 939.565 MeV$. \cite{pdg} \\
As for today, 28 total states are listed with PDG for $N^{*}$ which were a few before a decade as shown in \ref{tab:N-pdg}. ClAS experiment is focused to study nucleon structure through KY production \cite{carman}.  Unlike the ground state, excited state masses for both isospins are categorised under a common $N^{*}$, as it can be revealed by the relevant decay channel for the given mass\cite{hunt}. The N(1440) resonance in $J^{P} = \frac{1}{2}^{+}$, also known as the Roper resonance, is one of the most interesting states among the nucleon resonances \cite{fujii}. The Roper resonance
is lighter than the lowest negative-parity nucleon
excitations, i.e., N(1535) in  $J^{P} = \frac{1}{2}^{-}$ \cite{twanaka} and N(1520) in
$J^{P} = \frac{3}{2}^{-}$, which cannot be easily explained if one assumes
that the Roper resonance is a radial excitation of nucleon as a three-quark system. A promising physical interpretation is that the Roper resonance is the first radial excitation of nucleon but consists of a dressed-quark core augmented by a meson cloud. One of the most prominent examples of a baryon spectrum riddle is presented here. However, the present study is not able to comment on the nature of Roper resonance but the obtained result is well within the mass-range.
\begin{table}
	
	\caption{\label{tab:N-pdg} Experimental Status of all known $N^{*}$ \cite{pdg}}
	\centering
	\begin{tabular}{cc|cc|cc|cc}
		\hline
		**** & states & *** & states & ** & states & * states\\
		\hline
		N(1440) & $\frac{1}{2}^{+}$ &  N(1700) & $\frac{3}{2}^{-}$ & N(1860) & $\frac{5}{2}^{+}$ & N(2040) & $\frac{3}{2}^{+}$\\
		N(1520) & $\frac{3}{2}^{-}$ & N(1875) & $\frac{3}{2}^{-}$ & N(1990) & $\frac{7}{2}^{+}$ & \\
		N(1535) & $\frac{1}{2}^{-}$ & N(1880) & $\frac{1}{2}^{+}$ & N(2000) & $\frac{5}{2}^{+}$ & \\
		N(1650) & $\frac{1}{2}^{-}$ & N(2060) & $\frac{5}{2}^{-}$ & N(2300) & $\frac{1}{2}^{+}$ & \\
		N(1675) & $\frac{5}{2}^{-}$ & N(2100) & $\frac{1}{2}^{+}$ & N(2570) & $\frac{5}{2}^{-}$ & \\
		N(1680) & $\frac{5}{2}^{+}$ & N(2120) & $\frac{3}{2}^{-}$ & N(2700) & $\frac{13}{2}^{+}$ & \\
		N(1710) & $\frac{1}{2}^{+}$ & N(2600) & $\frac{11}{2}^{-}$ & & \\
		N(1720) & $\frac{3}{2}^{+}$ & & & & \\
		N(1895) & $\frac{1}{2}^{-}$ & & & &\\
		N(1900) & $\frac{3}{2}^{+}$ & & & & \\
		N(2190) & $\frac{7}{2}^{-}$ & & & & \\
		N(2220) & $\frac{9}{2}^{+}$ & & & & \\
		N(2250) & $\frac{9}{2}^{-}$ & & & & \\
		\hline
	\end{tabular}
\end{table}

\begin{table}
	
	\caption{\label{tab:N-screen} Resonance masses of $N^{*}$ using Screened potential (in MeV).}
	
	\begin{minipage}{0.48\textwidth}
		\centering
		\begin{tabular}{ccccc}
			\hline
			State & $J^{P}$ & $M_{scr}$ &  $M_{exp}$\\
			\hline
			1S & $\frac{1}{2}^{+}$ &939 & 938\\
			2S & $\frac{1}{2}^{+}$ & 1420 & 1440\\
			3S & $\frac{1}{2}^{+}$ & 1762 & 1710\\
			4S & $\frac{1}{2}^{+}$ & 2090 & 2040* \\
			5S & $\frac{1}{2}^{+}$ & 2422 & \\
			\hline
			$1^{2}P_{1/2}$ & $\frac{1}{2}^{-}$ & 1505 &  1535\\
			$1^{2}P_{3/2}$ & $\frac{3}{2}^{-}$ & 1493 & 1520\\
			$1^{4}P_{1/2}$ & $\frac{1}{2}^{-}$ & 1512 & 1650\\
			$1^{4}P_{3/2}$ & $\frac{3}{2}^{-}$ & 1499 & \\
			$1^{4}P_{5/2}$ & $\frac{5}{2}^{-}$ & 1482 & 1675\\
			\hline
			$2^{2}P_{1/2}$ & $\frac{1}{2}^{-}$ & 1882 & 1895 \\
			$2^{2}P_{3/2}$ & $\frac{3}{2}^{-}$ & 1868 & 1875 \\
			$2^{4}P_{1/2}$ & $\frac{1}{2}^{-}$ & 1890 \\
			$2^{4}P_{3/2}$ & $\frac{3}{2}^{-}$ & 1875 \\
			$2^{4}P_{5/2}$ & $\frac{5}{2}^{-}$ & 1856 \\
			\hline
			$3^{2}P_{1/2}$ & $\frac{1}{2}^{-}$ & 2286\\
			$3^{2}P_{3/2}$ & $\frac{3}{2}^{-}$ & 2270\\
			$3^{4}P_{1/2}$ & $\frac{1}{2}^{-}$ & 2294\\ 
			$3^{4}P_{3/2}$ & $\frac{3}{2}^{-}$ & 2278\\
			$3^{4}P_{5/2}$ & $\frac{5}{2}^{-}$ & 2257\\
			\hline
			$4^{2}P_{1/2}$ & $\frac{1}{2}^{-}$ & 2709\\
			$4^{2}P_{3/2}$ & $\frac{3}{2}^{-}$ & 2693 \\
			$4^{4}P_{1/2}$ & $\frac{1}{2}^{-}$ & 2717\\
			$4^{4}P_{3/2}$ & $\frac{3}{2}^{-}$ & 2701\\
			$4^{4}P_{5/2}$ & $\frac{5}{2}^{-}$ & 2679 \\
			\hline
			$1^{2}D_{3/2}$ & $\frac{3}{2}^{+}$ & 1816 &  1720\\
			$1^{2}D_{5/2}$ & $\frac{5}{2}^{+}$ & 1792 & 1680\\
			$1^{4}D_{1/2}$ & $\frac{1}{2}^{+}$ & 1843 & 1880\\
			$1^{4}D_{3/2}$ & $\frac{3}{2}^{+}$ & 1825 & 1900\\
			$1^{4}D_{5/2}$ & $\frac{5}{2}^{+}$ & 1801 & 1860\\
			$1^{4}D_{7/2}$ & $\frac{7}{2}^{+}$ & 1771 \\
			\hline
		\end{tabular}
	\end{minipage}
	\hfill
	\begin{minipage}{0.48\textwidth}
		\centering
		\begin{tabular}{ccccc}
			\hline
			State & $J^{P}$ & $M_{scr}$ &  $M_{exp}$\\
			\hline
			$2^{2}D_{3/2}$ & $\frac{3}{2}^{+}$ & 2215\\
			$2^{2}D_{5/2}$ & $\frac{5}{2}^{+}$ & 2190 \\
			$2^{4}D_{1/2}$ & $\frac{1}{2}^{+}$ & 2243\\
			$2^{4}D_{3/2}$ & $\frac{3}{2}^{+}$ & 2224\\
			$2^{4}D_{5/2}$ & $\frac{5}{2}^{+}$ & 2199\\
			$2^{4}D_{7/2}$ & $\frac{7}{2}^{+}$ & 2168 \\
			\hline
			$3^{2}D_{3/2}$ & $\frac{3}{2}^{+}$ & 2635\\
			$3^{2}D_{5/2}$ & $\frac{5}{2}^{+}$ & 2610\\
			$3^{4}D_{1/2}$ & $\frac{1}{2}^{+}$ & 2663\\
			$3^{4}D_{3/2}$ & $\frac{3}{2}^{+}$ & 2644\\
			$3^{4}D_{5/2}$ & $\frac{5}{2}^{+}$ & 2619\\
			$3^{4}D_{7/2}$ & $\frac{7}{2}^{+}$ & 2588 \\
			\hline
			$1^{2}F_{5/2}$ & $\frac{5}{2}^{-}$  & 2143 & 2060\\
			$1^{2}F_{7/2}$ & $\frac{7}{2}^{-}$ & 2107 & 2190 \\
			$1^{4}F_{3/2}$ & $\frac{3}{2}^{-}$ & 2183 & 2120\\
			$1^{4}F_{5/2}$ & $\frac{5}{2}^{-}$ & 2154 & \\
			$1^{4}F_{7/2}$ & $\frac{7}{2}^{-}$ & 2118 & 2190 \\
			$1^{4}F_{9/2}$ & $\frac{9}{2}^{-}$ & 2074 & 2250\\
			\hline
			$2^{2}F_{5/2}$ & $\frac{5}{2}^{-}$ & 2561 & 2570 \\
			$2^{2}F_{7/2}$ & $\frac{7}{2}^{-}$ & 2523 & \\
			$2^{4}F_{3/2}$ & $\frac{3}{2}^{-}$ & 2602\\
			$2^{4}F_{5/2}$ & $\frac{5}{2}^{-}$ & 2572\\
			$2^{4}F_{7/2}$ & $\frac{7}{2}^{-}$ & 2534\\
			$2^{4}F_{9/2}$ & $\frac{9}{2}^{-}$ & 2489\\
			\hline
			$1^{2}G_{7/2}$ & $\frac{7}{2}^{+}$ & 2487 & \\
			$1^{2}G_{9/2}$ & $\frac{9}{2}^{+}$ & 2433 & 2220\\
			$1^{4}G_{5/2}$ & $\frac{5}{2}^{+}$ & 2546\\
			$1^{4}G_{7/2}$ & $\frac{7}{2}^{+}$ & 2501\\
			$1^{4}G_{9/2}$ & $\frac{9}{2}^{+}$ & 2447\\
			$1^{4}G_{11/2}$ & $\frac{11}{2}^{+}$ & 2383\\
			\hline
			$1^{4}H_{11/2}$ & $\frac{11}{2}^{-}$ & 2786 & 2600\\
			\hline
		\end{tabular}
	\end{minipage}
\end{table}
The ground state parameters lead to the value of 938 MeV for P and 948 MeV for N, whereas earlier both the masses were 939 MeV such that excited states cannot be separated out. Table \ref{tab:N-screen} describes the obtained mass using the above phenomenological approach. The S-wave states are within a good range from those of Particle Data Group (PDG). The next four star status N(1440) with $J^{P}=\frac{1}{2}^{+}$ is the 2S state and present result 1420 is well within the PDG range and other approaches too. Similarly the 3S and 4S states are found to be with good agreement with experimental results. \\

In case of 1P states, the higher spin states have lower mass compared to their respective spin states for a given angular momentum. This has been observed has an intrinsic to hypercentral Constituent Quark Model (hCQM). 1P$\frac{1}{2}^{-}$ is 30 MeV less compared to 1535 MeV. All the spin states for 1D are experimentally established. The variation of splitting from $\frac{1}{2}$ to $\frac{7}{2}$ is around 80 MeV for the present masses. The very first state in negative parity is N(1520) with $J^{P}=\frac{3}{2}^{-}$ is reproduced as 1493. Also, this state is lower than its spin partner i.e. $J^{P}=\frac{1}{2}^{-}$ N(1535) is also consistent with our results as the model is predicting lower mass for higher spin state. Here, it is noteworthy that N(1650) $J^{P}=\frac{1}{2}^{-}$ doesn't appear in the present data.  N(1720)$\frac{3}{2}^{+}$ with four star label is 1816 in present results. 1D$\frac{1}{2}^{+}$ appears as a higher state than $\frac{3}{2}^{+}$ and $\frac{5}{2}^{+}$. Also, states in 2D are found to vary within 100 MeV difference compared to all. \\

All the negative parity states of 1F have been observed and assigned three and four star status. In the present results, higher spin state with $J^{P}=\frac{9}{2}^{-}$  is under-predicted compared to PDG. The two states which doesn't appear in earlier study have been calculated here. The N(2220) 1G $\frac{9}{2}^{+}$  obtained to be 2433 which is quite differing from PDG range. Also, the N(2600) is assigned 1H $\frac{11}{2}^{-}$ is over-predicted in the current results. Not all the models have resolved the hyperfine splitting of masses and also that hierarchy is not maintained in all the results. Recent studies have focused on N(1895) state and its decay to light hyperons $\gamma p \rightarrow K^{+}\Lambda$ \cite{khemchandani2022, martinez}. Out of the four states in the vicinity from 1875 to 1900, the negative parity states of our results are in accordance PDG range. But for positive parity states lying in D states cannot be precisely matched. This is true for other model comparisons as well.  \\
It is noteworthy that for low-lying states the masses with screened potential are in slightly increment which then falls with higher excited states. However, as the experimental masses fall within a range the predictions of screened and linear potential are not very far. The notable change comes into the picture with higher order correction terms.  \\

$\Delta$ baryon has played prime role towards the understanding of color quantum number. Even till date, $\Delta$ is important candidate not only in the field of high energy physics, but nuclear and astrophysical systems also \cite{marquez}. 
Over the course of many years, pion-nucleon decays and photoproduction decays have allowed for the observation of $\Delta$s. In the field of astrophysics,  $\Delta$ isobars are investigated under the quark-meson coupling model to determine whether if they could possibly be observed.   Incorporating the recent additions, 8 four star, 4 three star and many other experimental status have been explored with the values ranging from $J=\frac{1}{2}$ to $J=\frac{15}{2}$ and still many states are awaited of confirmation of existence as listed by Particle Data Group (PDG). \\
Also, $\Delta$ being the lightest member with the presence of electric quadrupole moment makes it interesting to dig deep into the shape and structure of the baryon \cite{ramalho}. Also, the MicroBooNE collaboration has recently reported $\Delta$(1232) radiative decay through neutrino induced neutral current \cite{microboone}. The pole positions for N and $\Delta$ resonances have been investigated through photoproduction of K$\Sigma$ from coupled-channel study \cite{ronchen}. \\

\begin{table}
	\caption{\label{tab:Delta-screen}Resonance Masses of $\Delta$ baryon using Screened Potential (in MeV)}
	
	\begin{minipage}{0.48\textwidth}
		\centering
		\begin{tabular}{ccccc}
			\hline
			State & $J^{P}$ & $M_{scr}$ & $M_{exp}$ \\
			\hline
			1S & $\frac{3}{2}^{+}$ & 1232 & 1232\\
			2S & $\frac{3}{2}^{+}$ & 1602 & 1600\\
			3S & $\frac{3}{2}^{+}$  & 1900 & 1910\\
			4S & $\frac{3}{2}^{+}$ & 2200 & \\ 
			5S & $\frac{3}{2}^{+}$ & 2499 & \\
			\hline
			$1^{2}P_{1/2}$ & $\frac{1}{2}^{-}$ & 1556 & 1620\\
			$1^{2}P_{3/2}$ & $\frac{3}{2}^{-}$ & 1542 & 1700\\
			$1^{4}P_{1/2}$ & $\frac{1}{2}^{-}$ & 1563 &\\
			$1^{4}P_{3/2}$ & $\frac{3}{2}^{-}$ & 1549 & \\
			$1^{4}P_{5/2}$ & $\frac{5}{2}^{-}$ & 1530 & \\
			\hline
			$2^{2}P_{1/2}$ & $\frac{1}{2}^{-}$ & 1911 & 1900\\
			$2^{2}P_{3/2}$ & $\frac{3}{2}^{-}$ & 1895 & \\
			$2^{4}P_{1/2}$ & $\frac{1}{2}^{-}$ & 1919 & \\
			$2^{4}P_{3/2}$ & $\frac{3}{2}^{-}$ & 1903 & 1940\\
			$2^{4}P_{5/2}$ & $\frac{5}{2}^{-}$ & 1882 & 1930\\
			\hline
			$3^{2}P_{1/2}$ & $\frac{1}{2}^{-}$ & 2272 & \\
			$3^{2}P_{3/2}$ & $\frac{3}{2}^{-}$ & 2257\\
			$3^{4}P_{1/2}$ & $\frac{1}{2}^{-}$ & 2280\\
			$3^{4}P_{3/2}$ & $\frac{3}{2}^{-}$ & 2265\\
			$3^{4}P_{5/2}$ & $\frac{5}{2}^{-}$ & 2243 & 2350*\\
			\hline
			$4^{2}P_{1/2}$ & $\frac{1}{2}^{-}$ & 2638\\
			$4^{2}P_{3/2}$ & $\frac{3}{2}^{-}$ & 2623\\
			$4^{4}P_{1/2}$ & $\frac{1}{2}^{-}$ & 2646\\
			$4^{4}P_{3/2}$ & $\frac{3}{2}^{-}$ & 2631\\
			$4^{4}P_{5/2}$ & $\frac{5}{2}^{-}$ & 2610\\
			\hline
			
			$1^{2}D_{3/2}$ & $\frac{3}{2}^{+}$ & 1841 & \\
			$1^{2}D_{5/2}$ & $\frac{5}{2}^{+}$ & 1817 & 1905\\
			$1^{4}D_{1/2}$ & $\frac{1}{2}^{+}$ & 1867\\
			$1^{4}D_{3/2}$ & $\frac{3}{2}^{+}$ & 1850 & 1920\\
			$1^{4}D_{5/2}$ & $\frac{5}{2}^{+}$ & 1826\\
			$1^{4}D_{7/2}$ & $\frac{7}{2}^{+}$ & 1797 & 1950\\
			\hline
			$2^{2}D_{3/2}$ & $\frac{3}{2}^{+}$ & 2202 \\
			$2^{2}D_{5/2}$ & $\frac{5}{2}^{+}$ & 2178\\
			\hline
		\end{tabular}
	\end{minipage}
	\hfill
	\begin{minipage}{0.48\textwidth}
		\centering
		\begin{tabular}{ccccc}
			\hline
			State & $J^{P}$ & $M_{scr}$ & $M_{exp}$ \\
			\hline
			$2^{4}D_{1/2}$ & $\frac{1}{2}^{+}$ & 2230 & \\
			$2^{4}D_{3/2}$ & $\frac{3}{2}^{+}$ & 2211\\
			$2^{4}D_{5/2}$ & $\frac{5}{2}^{+}$ & 2187\\
			$2^{4}D_{7/2}$ & $\frac{7}{2}^{+}$ & 2156\\
			\hline
			$3^{2}D_{3/2}$ & $\frac{3}{2}^{+}$ & 2566\\
			$3^{2}D_{5/2}$ & $\frac{5}{2}^{+}$ & 2543\\
			$3^{4}D_{1/2}$ & $\frac{1}{2}^{+}$ & 2593 \\
			$3^{4}D_{3/2}$ & $\frac{3}{2}^{+}$ & 2575\\
			$3^{4}D_{5/2}$ & $\frac{5}{2}^{+}$ & 2552\\
			$3^{4}D_{7/2}$ & $\frac{7}{2}^{+}$ & 2523\\
			\hline
			$4^{2}D_{3/2}$ & $\frac{3}{2}^{+}$ & 2934\\
			$4^{2}D_{5/2}$ & $\frac{5}{2}^{+}$ & 2912\\
			$4^{4}D_{1/2}$ & $\frac{1}{2}^{+}$  & 2959\\
			$4^{4}D_{3/2}$ & $\frac{3}{2}^{+}$ & 2942\\
			$4^{4}D_{5/2}$ & $\frac{5}{2}^{+}$ & 2921\\
			$4^{4}D_{7/2}$ & $\frac{7}{2}^{+}$ & 2893\\
			\hline
			$1^{2}F_{5/2}$ & $\frac{5}{2}^{-}$ & 2131 & \\
			$1^{2}F_{7/2}$ & $\frac{7}{2}^{-}$ & 2095 & 2200\\
			$1^{4}F_{3/2}$ & $\frac{3}{2}^{-}$ & 2170 & \\
			$1^{4}F_{5/2}$ & $\frac{5}{2}^{-}$ & 2141 & \\
			$1^{4}F_{7/2}$ & $\frac{7}{2}^{-}$ & 2106 & \\
			$1^{4}F_{9/2}$ & $\frac{9}{2}^{-}$ & 2063 & \\
			\hline
			
			$1^{2}G_{7/2}$ & $\frac{7}{2}^{+}$ & 2420 & \\
			$1^{2}G_{9/2}$ & $\frac{9}{2}^{+}$ & 2375 & 2300\\
			$1^{4}G_{5/2}$ & $\frac{5}{2}^{+}$ & 2468 & \\
			$1^{4}G_{7/2}$ & $\frac{7}{2}^{+}$ & 2431 & 2390\\
			$1^{4}G_{9/2}$ & $\frac{9}{2}^{+}$ & 2387 & \\
			$1^{4}G_{11/2}$ & $\frac{11}{2}^{+}$ & 2335 & 2420\\
			\hline
			$1^{2}H_{9/2}$ & $\frac{9}{2}^{-}$ & 2719 & \\
			$1^{2}H_{11/2}$ & $\frac{11}{2}^{-}$ & 2657 & \\
			$1^{4}H_{7/2}$ & $\frac{7}{2}^{-}$ & 2786 & \\
			$1^{4}H_{9/2}$ & $\frac{9}{2}^{-}$ & 2732 & \\
			$1^{4}H_{11/2}$ & $\frac{11}{2}^{-}$ & 2670 & \\
			$1^{4}H_{13/2}$ & $\frac{13}{2}^{-}$ & 2600 & 2750\\
			\hline
		\end{tabular}
	\end{minipage}
\end{table}

Similar to N, the mass spectra of $\Delta$ has been tabulated in table \ref{tab:Delta-screen}. The S-wave mass predictions are very well matching with experimental data. The 1P(1556) state is 70 MeV lower compared to 1620 MeV. However, the 1P$\frac{3}{2}^{-}$ state is 150 MeV under-predicted. Few states in 2P show good agreement. The 1D states are under-predicted compared to PDG. The 1G $\frac{9}{2}^{+}$ is in accordance with PDG value of 2300 MeV by a difference of 75 MeV. 1H $\frac{13}{2}^{-}$ (2750) is predicted to be 2600 MeV in the present work. \\ 

 As the screened potential has been observed in heavy quark systems, our primary goal here to check if similar kind of effects apply to light quark systems. Briefly concluding that screening effect in light systems is notable at higher mass scale but not too suppressed as in the case of heavy systems. Another important aspect is the higher order corrections do not resolve the spin structure in case of screened potential.

\section{Regge Trajectory}
One of the helpful tools in spectroscopic research has been Regge trajectories. Based on calculated data, figures depict a plot of total angular momentum J and principle quantum number n versus the square of resonance mass $M^2$. Many studies have found that the theoretical and experimental data are consistent with the non-intersecting and linearly fitted lines \cite{jpac}. A correct spin-parity assignment for a state can perhaps be predicted using these plots.
\begin{subequations}
	\begin{align}
		J = aM^{2} + a_{0} \\
		n = b M^{2} + b_{0}
	\end{align}
\end{subequations}
\begin{figure}
	\centering
	\includegraphics[scale=0.4]{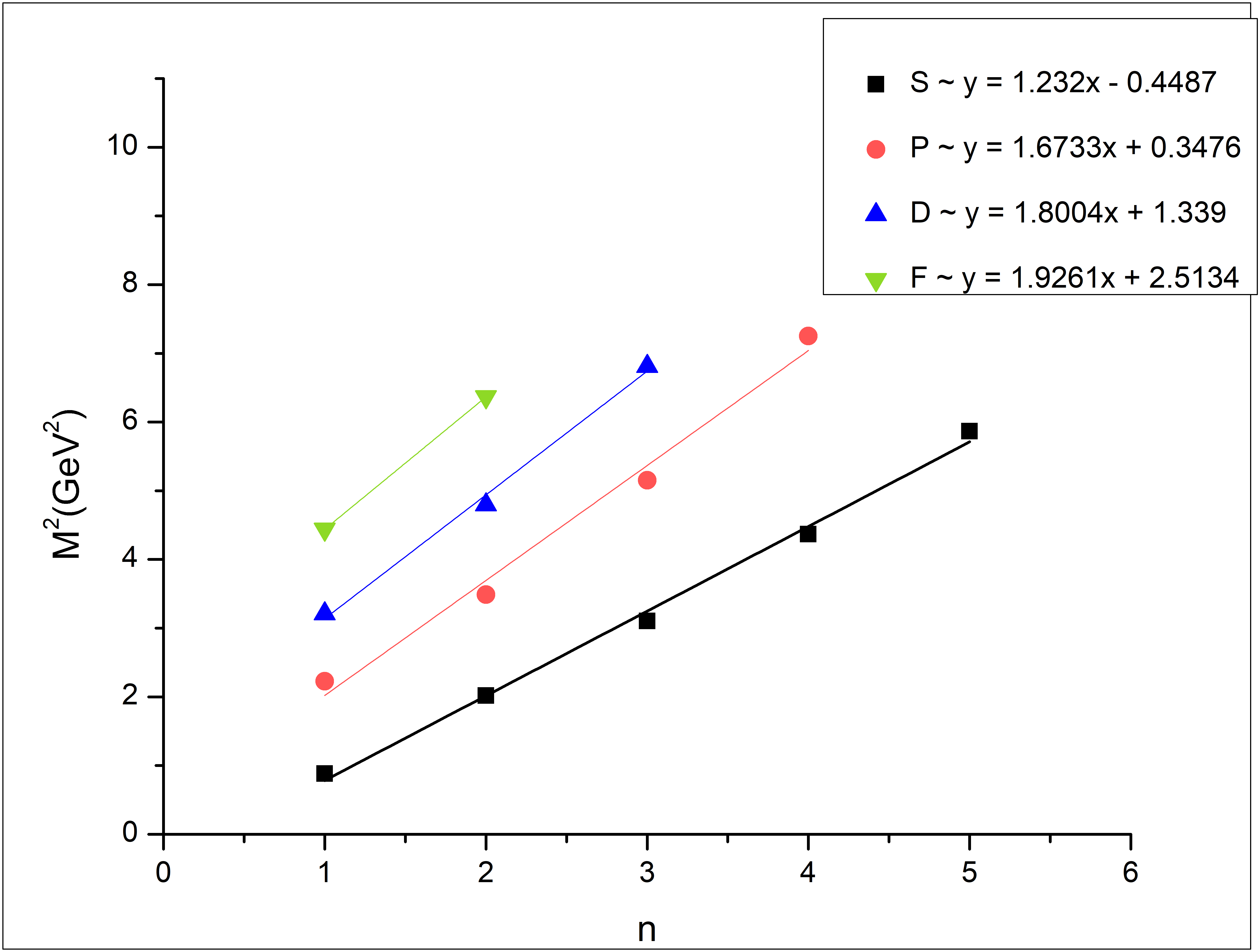}
	\caption{\label{fig:n-n-s} (n,$M^{2}$) Regge trajectory for $N$ baryon for screened potential.}
\end{figure}

\begin{figure}
	\centering
	\includegraphics[scale=0.4]{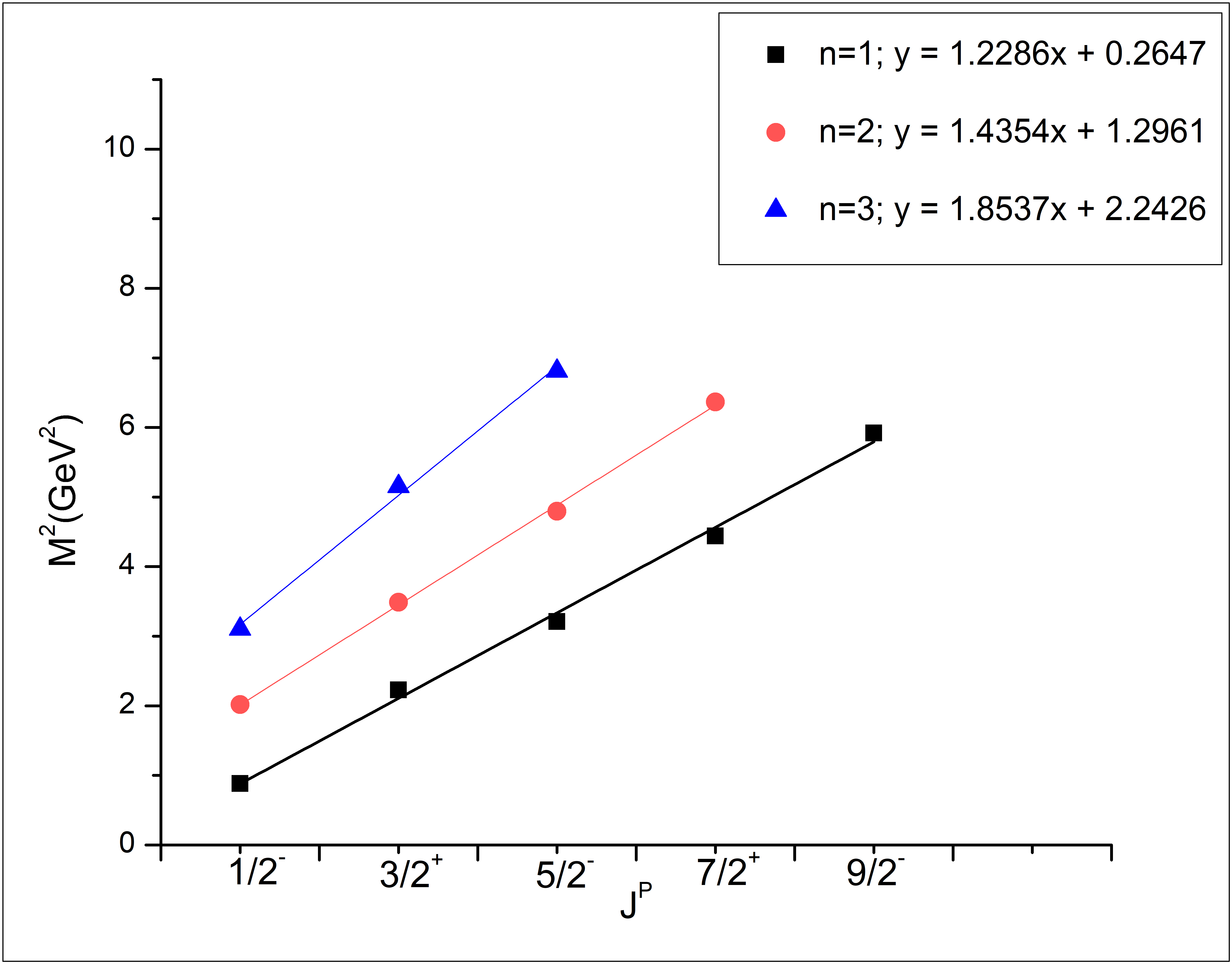}
	\caption{\label{fig:n-j1-s} (J,$M^{2}$) Regge trajectory for N baryon for screened potential.}
\end{figure}
\begin{figure}
	\centering
	\includegraphics[scale=0.4]{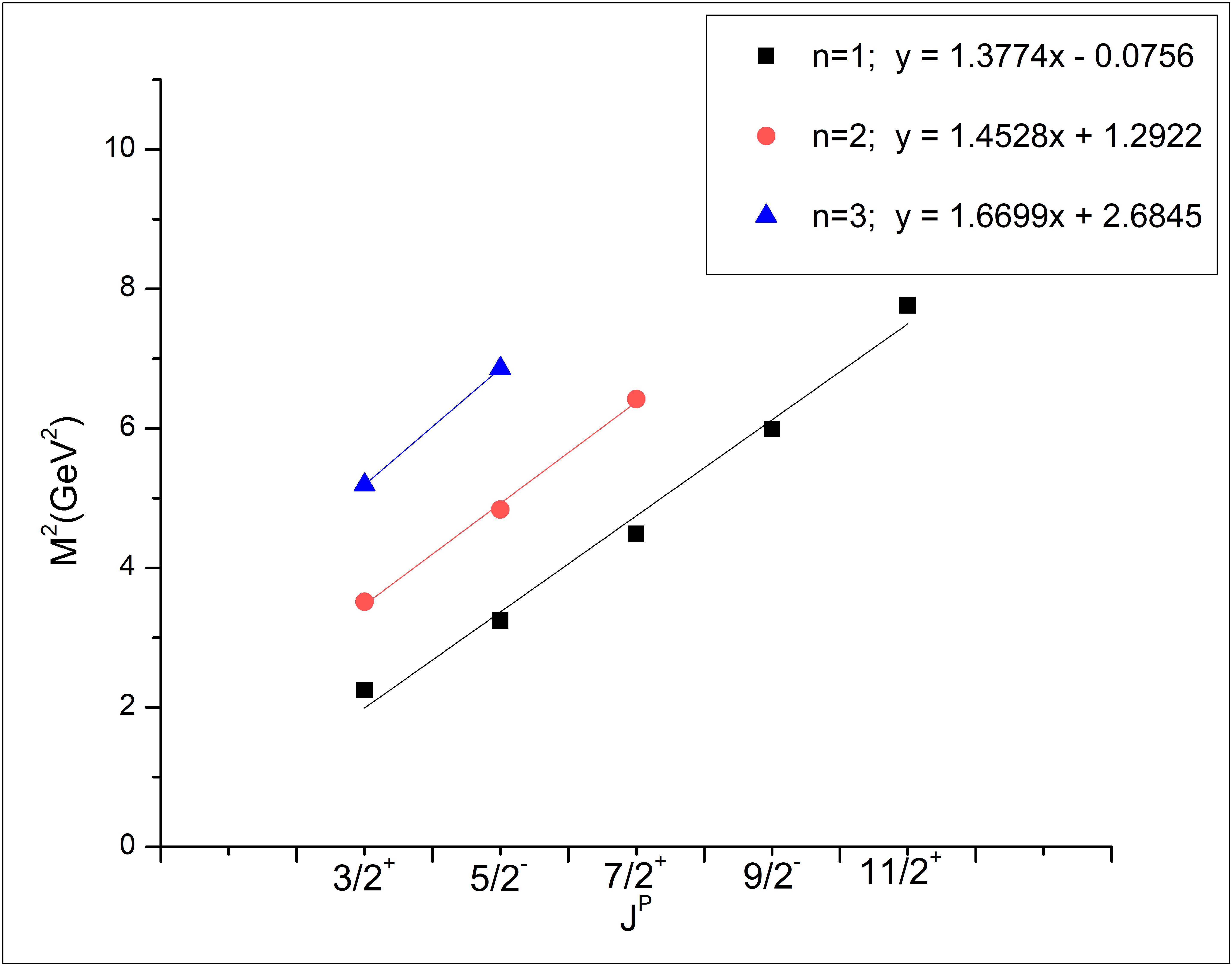}
	\caption{\label{fig:n-j3-s} (J,$M^{2}$) Regge trajectory for N baryon for screened potential.}
\end{figure}

\begin{figure}
	\centering
	\includegraphics[scale=0.4]{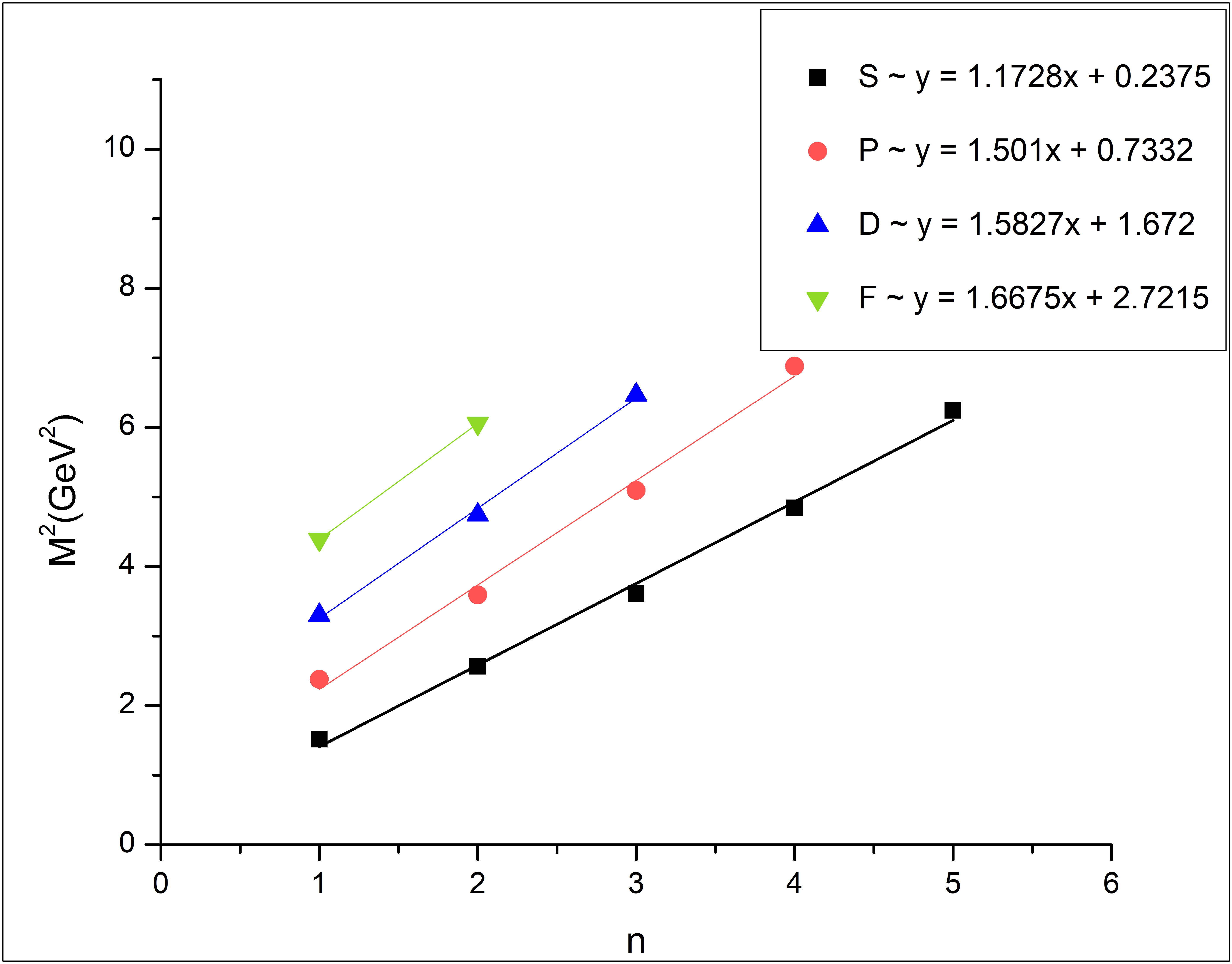}
	\caption{\label{fig:delta-n-s} (n,$M^{2}$) Regge trajectory for $\Delta$ states for screened potential}
\end{figure}

\begin{figure}
	\centering
	\includegraphics[scale=0.4]{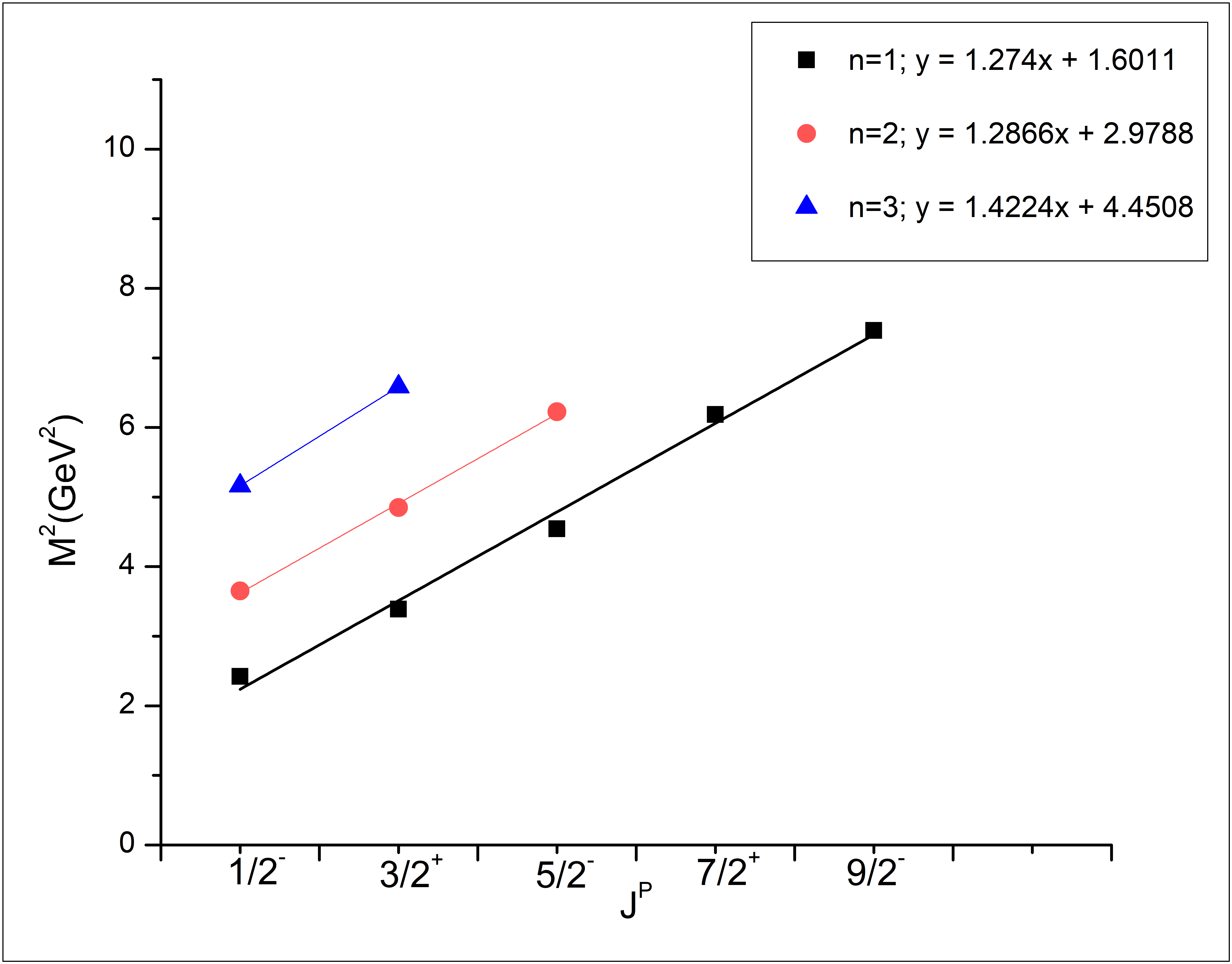}
	\caption{\label{fig:delta-j1-s} (J,$M^{2}$) Regge trajectory for $\Delta$ states for screened potential}
\end{figure}

\begin{figure}
	\centering
	\includegraphics[scale=0.4]{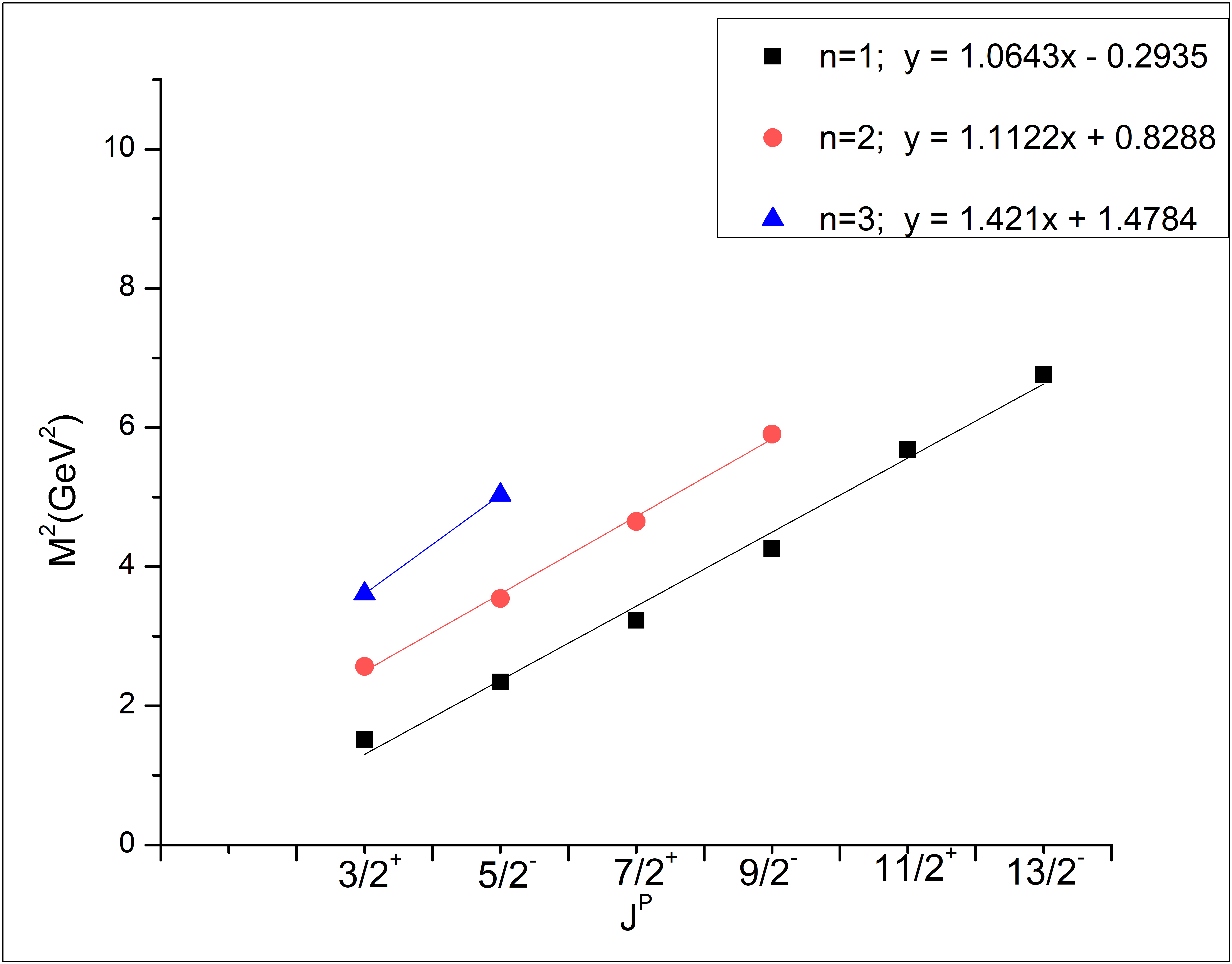}
	\caption{\label{fig:delta-j3-s} (J,$M^{2}$) Regge trajectory for $\Delta$ states for screened potential}
\end{figure}

The Regge trajectories have been observed to be following linear nature as expected. However, it is observed that all the lines are not equidistant. The slopes and intercepts have been mentioned in the graphs itself. 

\section{Decay}
In the case of nucleons, including $\Delta$, the prominent decay channel has been observed to be either $N*$ or pion, depending on the charge of the respective parent \cite{elsa}. In addition to other constants, the transition couplings of vector mesons have been obtained thanks to the work of Riska and colleagues \cite{riska}. Lagrangian densities for the transition couplings involving generalized Rarita-Schwinger vector spinors are used to define the coupling constants of the vector-meson transition to nucleon resonances. The transition coupling constants can be written in terms of the corresponding vector-meson coupling constants to the nucleons by comparing the matrix elements of these Lagrangians to the corresponding matrix elements in the quark model. The latter are calculated using phenomenological boson exchange interaction models and fitted to nucleon-nucleon scattering data, albeit with large uncertainty margins. P- and D-shell and excited S-shell states can be related to the ground state through these expressions involving SU(2) Clebsch-Gordan coefficients and orbital matrix elements of quark wave functions. The latter are dependent on the Hamiltonian model of the three-quark system. Here we use a straightforward model for covariant harmonic oscillators where the confining interaction is linear and the hyperfine interaction depends on the flavour and spin of the particle. In the work that is being presented here, the constants and decay widths that have been supplied by the Particle Data group have been used to determine the decay width of certain resonance masses that have already been established.\\

For $\Delta(1232)$, $\Delta(1600)$ decay to $N\pi$,
\begin{equation}
	\Gamma = \frac{1}{3}\frac{f^{2}}{4\pi}\frac{E^{'}+m_N}{m_{\Delta}}\frac{k^{3}}{m_{\pi}^{2}}
\end{equation}
where, $E^{'}$ is the energy of the final nucleon and k is pion momentum.
\begin{equation}
	E^{'} = \frac{m^{*2}-m_{\pi}^{2}+m_N^{2}}{2m^{*}}
\end{equation}
\begin{equation}
	k = \frac{\sqrt{[m^{*2}-(m_N + m_{\pi})^{2}][m^{*2}-(m_N - m_{\pi})^{2}]}}{2m^{*}}
\end{equation}
Here $m^{*}$ is resonance mass calculated using above model, $m_{N}$ is nucleon mass 939 MeV and $m_{\pi}$ is pion mass 139 MeV. 
For $N(1535)$, $N(1650)$ and $\Delta(1620)$ decaying to $ N\pi$,
\begin{equation}
	\Gamma = \frac{f^{2}}{4\pi}\frac{E^{'}+m_N}{m^{*}}\frac{k}{m_{\pi}^{2}}(m^{*}-m_{N})^{2}
\end{equation}

For $N(1520)$, $N(1700)$ and $\Delta(1700)$ decaying to $ N\pi$,
\begin{equation}
	\Gamma = \frac{1}{3}\frac{f^{2}}{4\pi}\frac{E^{'}-m_N}{m_{\Delta}}\frac{k^{3}}{m_{\pi}^{2}}
\end{equation}

\begin{table}
	\centering
	\caption{Pion decay widths $\Gamma$ for N and $\Delta$ channels}
	\label{tab:decay}
	\begin{tabular}{cccccc}
		\hline
		Decay & Mass & $\Gamma$(Present work) & \cite{hunt} & \cite{ardnt} &  \\
		\hline
	$\Delta$(1232)$\frac{3}{2}^{+}$ & 1232 & 119.855 & & 118.7(0.6)\\
	$\Delta$(1600)$\frac{3}{2}^{+}$ & 1602 & 60.586 & 34(8) \\
	N(1440)$\frac{1}{2}^{+}$ & 1420 & 209.962 & 153 & 284(18) \\
	N(1535)$\frac{1}{2}^{-}$ & 1505 & 99.961 & 62(3) & 188.4(3.8)\\
	N(1650)$\frac{1}{2}^{-}$ & 1512 & 76.5586 & 86(6) & 115.4(2.8) \\
	$\Delta$(1620)$\frac{1}{2}^{-}$ & 1556 & 111.779 & 26(2) & 146.9(1.9)\\
	N(1520)$\frac{3}{2}^{-}$ & 1493 & 17.87 & 71(2) & 103.6(0.4)\\
	$\Delta$(1700)$\frac{3}{2}^{-}$ & 1542 & 17.85 & 3 & 375.5\\
	N(1675)$\frac{5}{2}^{-}$ & 1482 & 16.29 & 53 & 146.5\\
	N(1720)$\frac{3}{2}^{+}$ & 1816 & 27.80 &  41 & 210(22) \\
	N(1680)$\frac{5}{2}^{+}$ & 1792 & 178.76 & 84 \\
	\hline

	\end{tabular}
\end{table}

The values for decay constant f varies with each decay channel. These values are descriptively studied by Riska and Brown. Table \ref{tab:decay} shows decay channels of N and $\Delta$ baryon with respective decay widths obtained for our predicted resonance masses. The comparison of decay widths with a recent experimental findings as elaborated by Hunt et al. \cite{hunt}. Our results are in good agreement in few channels. Also, we have compared with another partial wave analysis done by Ardnt et al. \cite{ardnt}.

\section{Conclusion}
The present work summarizes the effect of screened type potential under hypercentral Constituent Quark Model (hCQM) for N and $\Delta$ baryons. So far, linear potential has been applied to light spectrum, whereas screened potential provided reasonable results for heavy quark systems. The screening parameter plays a role in determining the spin-split and mass at higher angular momentum states. The obtained masses have been comparable with the basis of experimental known values with different star status. The masses of higher spin state for a given L value, decreases in hierarchy. The hyperfine splitting is observed to be less with screened potential than those of linear one. \\
The Regge trajectories show linear trend for all natural and unnatural parity points. The strong decay widths for pion channel are also calculated in the present study. The ongoing and upcoming experimental facilities namely HADES \cite{hades}, PANDA\cite{panda2,panda3,panda4,panda5,panda6,panda7}, shall be providing us with new insights in the understanding of light, strange baryons. 

\section*{Acknowledgment} 
Ms.Chandni Menapara acknowledges the support for pursuing this work under DST-INSPIRE Fellowship Scheme.

\bibliographystyle{natbib}

\end{document}